\documentclass[twocolumn,showpacs,preprintnumbers,amsmath,amssymb]{revtex4}

\usepackage{graphicx}
\usepackage{dcolumn}
\usepackage{bm}

\begin{document}

\preprint{APS}

\title{Hyperons in the relativistic mean-field approach to asymmetric nuclear matter}

\author{J. Kotuli\v{c} Bunta}
\email{juraj.bunta@savba.sk}

\author{\v{S}tefan Gmuca}
\email{gmuca@savba.sk}

\affiliation{Institute of Physics, Slovak Academy of Sciences\\
Dubravska cesta 9, SK-842 28 Bratislava, Slovak Republic}

\date{\today}

\begin{abstract}

Relativistic mean-field theory with $\delta$ meson, nonlinear
isoscalar self-interactions and isoscalar-isovector cross interaction
terms with parametrizations obtained to reproduce
Dirac-Brueckner-Hartree-Fock calculations for nuclear matter is
used to study asymmetric nuclear matter properties in
$\beta$-equilibrium, including hyperon degrees of freedom and
(hidden) strange mesons. Influence of cross interaction on
composition of hyperon matter and electron chemical potential is
examined. Softening of nuclear equation of state by the cross
interactions results in lowering of hyperonization, although
simultaneously enhancing a hyperon-induced decrease of the
electron chemical potential, thus indicating further shift of a
kaon condensate occurence to higher densities.
\end{abstract}



\maketitle

\section{\label{sec:level1}Introduction}

Nuclear matter in $\beta$-equilibrium is one of the main objects
of interest of nuclear physics with direct astrophysical
implications.

Several theoretical models were developed for describing the
nuclear matter. A direct connection between the fundamental
nucleon-nucleon interaction and nuclear matter properties is
provided by the Dirac-Brueckner-Hartree-Fock (DBHF) theory
\cite{BM84,HM87,JL98} which successfully describes properties of
symmetric and also asymmetric nuclear matter. However, this
sophisticated approach cannot be applied to finite nuclei
calculations yet. Thus, other approaches are still relevant. Due
to its elegant framework, the relativistic mean-field (RMF) theory
\cite{SW86,SW97,B91} has been widely and successfully applied to
wide range of finite nuclei. In order to eliminate
phenomenological nature of RMF theory the free parameters can be
fitted to the more fundamental DBHF approach \cite{G92}, thus
obtaining effective DBHF parametrization applicable also to finite
nuclei. An alternative way is using the density dependent
relativistic hadron theory \cite{FL95}.

In recent years a significant progress in development of
experimental techniques and devices was achieved. Modern
accelerators enabled study of high energy ion collisions producing
nuclei with high isospin asymmetry and nuclear densities higher
than those occurring in normal nuclei. Therefore, a special
attention was paid to a proper asymmetry description, e.g., an
enhancement of isovector meson sector by $\delta$ meson
introduction was performed \cite{KK97}. However, better
improvement of density dependence of symmetry energy was achieved
by including of the scalar-vector cross interaction (VCI) term
\cite{MS96}.

In addition to nuclear physics research also astrophysical
phenomena are starting to provide relevant data with an accuracy
needed to constrain the equation of state (EOS) of nuclear matter.
Neutron star masses and radii may be able to provide such
constraints, due to their essential dependence on EOS, and many
works deal with such calculations using DBHF approach
\cite{MPA88,HWW94,EHO94}, RMF theory \cite{Gl87,BM01}, or density
dependent hadron field theory \cite{HKL02}.

Since in the neutron star interior we expect densities of several
times of the nuclear saturation density, extrapolation of models
to high density region is inevitable. Classical view of matter as
consisting of protons, neutrons and electrons is thus insufficient
and more realistic composition is needed. At higher densities the
$K^{-}$-condensate \cite{KN86}, quark deconfinement \cite{CP75}
and/or hyperons \cite{MHB03} are possible to appear.

After theoretical suggestion of the hyperon \cite{AS60} and
hypernuclei \cite{Ty76} appearance in high density nuclear matter
many works paid attention to them, e.g.,
\cite{Gl85,WW89,RSM90,GPS95}, and experimental effort has started
to obtain empirical data \cite{BI89}. However, hyperon-meson
coupling constants still remain uncertain. A partial exception is
the $\Lambda$ hyperon where some constraint of its bindings is
possible \cite{MDG88}. Even poorer knowledge we have about the
$\Sigma^{-}$-nucleon interaction. There is an absence of bound
$\Sigma^{-}$ hypernuclear states \cite{Ba99} which might be a
support for a high density repulsion of $\Sigma^{-}$-nucleon force
as indicated by extrapolated atomic data \cite{FGB94}. However,
this is still an open question. In any case, it seems to have only
a weak effect on bulk properties of hyperon matter
\cite{Gl01,BLC99}, due to similar influence of the heavier
$\Xi^{-}$ hyperon, which (same as the $\Lambda$ hyperon) feels
attractive potential \cite{Kh00}. Previous deducing of the
$\Lambda\Lambda$ interaction from older double $\Lambda$
hypernuclei data suggested a highly attractive potential
\cite{FG02}, while the very recent experimental observation of a
$^{6}_{\Lambda\Lambda}$He double hypernucleus demonstrated
\cite{Takahashi01} that $\Lambda\Lambda$ interaction is only
weakly attractive. With respect to all these facts, we used
hyperon-nucleon and hyperon-hyperon coupling constants derived
from SU(6) quark model for vector and isovector meson sector.
Hyperon couplings to scalar mesons were fitted to reproduce
experimental data for hypernuclei and double hypernuclei (see the
next section for more details).

Recently, the RMF parametrization of DBHF results for nuclear
matter, including isovector scalar $\delta$ meson as well as
scalar-vector cross interactions \cite{KBG03}, was obtained. The
results indicate a strong influence of cross interaction on the
density dependence of symmetry energy. The goal of this paper is
an application of this parametrizations also to asymmetric nuclear
matter with hyperons, and thus to study the effect of additional
degrees of freedom on a composition of dense matter. The
theoretical framework described in the Section II. is followed by
presentation and discussion of the obtained results in the Section
III. Conclusions are summarized in the Section IV.

\section{The Model}

We start with the Lagrangian density that introduces baryon field
$\psi_{B}$, lepton fields $\psi_{e^{-}}$ and $\psi_{\mu^{-}}$,
isoscalar-scalar meson field $\sigma$, isoscalar-vector meson
field $\omega$, isovector-vector meson field $\bm{\rho}$,
isovector-scalar meson field $\bm{\delta}$ (pion field does not
contribute because of its pseudoscalar nature, and nuclear matter
is parity invariant). Hidden strange scalar meson $\sigma^{*}$ and
vector meson $\phi$ are introduced in order to better simulate
hyperon-hyperon attraction observed in $\Lambda\Lambda$
hypernuclei \footnote{The original motivation of strange meson
introduction was the simulation of strong YY interaction. In spite
of the recent experimental proof of a weak $\Lambda\Lambda$
interaction in $^{6}_{\Lambda\Lambda}$He \cite{Takahashi01} there
still remain reasons for their incorporation \cite{SSL03}.}. Thus,
the Lagrangian density takes a form,

\begin{center}
\begin{eqnarray}
&{\cal{L}}(\psi_{B,e^{-},\mu^{-}},\sigma,\omega,\bm{\rho},\bm{\delta},\sigma^{*},\phi)
\nonumber\\
&=\sum_{B}\bar{\psi}_{B}
[\bm{\gamma}_{\mu}(i\partial^{\mu}-g_{\omega}\omega^{\mu}
-g_{\rho}\bm{\rho}_{\mu}\bar{\psi}\gamma^{\mu}\bm{\tau} -g_{\phi
B}\phi^{\mu})
\nonumber\\
&-(M-g_{\sigma}\sigma-g_{\delta}\bm{\delta}\bar{\psi}\bm{\tau}\psi
-g_{\sigma^{*} B}\sigma^{*})]\psi_{B}
\nonumber\\
&+\frac{1}{2}(\partial_{\mu}\sigma\partial^{\mu}\sigma-m_{\sigma}^{2}\sigma^{2})
-\frac{1}{3}b_{\sigma}M{(g_{\sigma}\sigma)}^{3}-\frac{1}{4}c_{\sigma}
{(g_{\sigma}\sigma)}^{4}
\nonumber\\
&\label{lagrangian}
-\frac{1}{4}\omega_{\mu\nu}\omega^{\mu\nu}+\frac{1}{2}m_{\omega}^{2}\omega_{\mu}
\omega^{\mu}
+\frac{1}{4}c_{\omega}{(g_{\omega}^{2}\omega_{\mu}\omega^{\mu})}^{2}
\nonumber\\
&+\frac{1}{2}(\partial_{\mu}\bm{\delta}\partial^{\mu}\bm{\delta}
-m_{\delta}^{2}{\bm{\delta}}^{2})
+\frac{1}{2}m_{\rho}^{2}\bm{\rho}_{\mu}.\bm{\rho}^{\mu}
-\frac{1}{4}\bm{\rho}_{\mu\nu}.\bm{\rho}^{\mu\nu}
\nonumber\\
&+\frac{1}{2}\Lambda_{V}(g_{\rho}^{2}\bm{\rho_{\mu}}.\bm{\rho}^{\mu})
(g_{\omega}^{2}\omega_{\mu}\omega^{\mu})
\nonumber\\
&+\frac{1}{2}(\partial_{\mu}\sigma^{*}\partial^{\mu}\sigma^{*}-
m_{\sigma^{*}}^{2}{\sigma^{*}}^{2})
+\frac{1}{2}{m_{\phi}^{2}}\phi_{\mu}\phi^{\mu}
-\frac{1}{4}\phi_{\mu\nu}\phi^{\mu\nu}
\nonumber\\
&+\sum_{e,\mu}\bar{\psi}_{e,\mu}(i\gamma_{\mu}\partial^{\mu}-m_{e,\mu})\psi_{e,\mu}
 \;
 \end{eqnarray}
\end{center}
where antisymmetric field tensors are given by

\begin{eqnarray*}
\omega_{\mu\nu}\equiv\partial_{\nu}\omega_{\mu}-\partial_{\mu}\omega_{\nu}
\;,
\\
\bm{\rho}_{\mu\nu}\equiv\partial_{\nu}\bm{\rho}_{\mu}
-\partial_{\mu}\bm{\rho}_{\nu}\;,\\
{\phi}_{\mu\nu}\equiv\partial_{\nu}\phi_{\mu}
-\partial_{\mu}\phi_{\nu} \;,
\end{eqnarray*}
and the symbols used have the same meaning as in Ref.
\cite{KBG03}.

The nucleon-meson coupling constants as well as isoscalar
self-interaction couplings and the vector cross interaction
coupling were taken from our previous work \cite{KBG03}. Coupling
constants of hyperons to the vector and isovector mesons as well
as vector strange mesons are derived from the SU(6) simple quark
model and are of the following form \cite{SDG94}:

\begin{eqnarray}\label{quarkmodel}
&\frac{1}{3}g_{\omega
N}=\frac{1}{2}g_{\omega\Lambda}=\frac{1}{2}g_{\omega\Sigma}\;,
\nonumber\\
&g_{\rho N}=\frac{1}{2}g_{\rho\Sigma}\enspace,\enspace
g_{\rho\Lambda}=0\;,
\nonumber\\
&g_{\delta N}=\frac{1}{2}g_{\delta\Sigma}\enspace,\enspace
g_{\delta\Lambda}=0\;,
\nonumber\\
& g_{\phi\Lambda}=g_{\phi\Sigma}=-\frac{\sqrt{2}}{3}g_{\omega N
}\enspace,\enspace g_{\phi N}=0\;.
\end{eqnarray}

Remaining couplings of hyperons to scalar mesons are adjusted to
reproduce the hypernuclear potentials in saturated nuclear matter.
A recent analysis \cite{MFG95} showed that $\Sigma^{-}$ can feel
repulsion in nuclear matter, thus leading to a strong suppression
of its abundance in $\beta$-stable hyperon matter. However, for
our purposes it is of a little importance because it has only a
minor effect on bulk hyperon matter properties, as was shown in
Ref. \cite{Gl01}. The matter is dominated by nucleons until high
densities where universal short-range forces are expected to take
precedence over the specific baryon identities \cite{BLC99}.
Therefore, we have taken into account the following recent
potential depths \cite{MDG88,SBG00}:

\begin{equation}\label{hypwell}
U_{\Lambda}^{(N)}=-28 \;\textrm{MeV} \;,\enspace
U_{\Sigma}^{(N)}=+20 \;\textrm{MeV}\;.
\end{equation}

Finally, the hyperon couplings to scalar strange mesons are fixed
by potential well of the $\Lambda$-hyperon in $\Lambda$-hyperonic
matter deduced from the recent double-$\Lambda$ hypernuclear data
\cite{Takahashi01}, as estimated by Song \cite{SSL03}.

\begin{equation}\label{doublehypwell}
U_{\Lambda}^{(\Lambda)}=-5 \;\textrm{MeV}\;.
\end{equation}

Application of Euler-Lagrange equations to the Lagrangian
(\ref{lagrangian}) leads to equations of motion. Baryons fulfill
the Dirac equation,

\begin{eqnarray} \label{dirac}
\sum_{B}\big[\bm{\gamma}_{\mu}(i\partial^{\mu}-g_{\omega}\omega^{\mu}
-g_{\rho}\bm{\rho}_{\mu}.\bm{\tau}-g_{\phi B}\bm{\phi}^{\mu})
\nonumber\\
-(M-g_{\sigma}\sigma-g_{\delta}\bm{\delta}.\bm{\tau}-g_{\sigma^{*}
B}\sigma^{*})\big]\psi_{B}=0\;,
\end{eqnarray}
while isoscalar $\sigma$, $\omega$, isovector $\bm{\delta}$,
$\bm{\rho}$ and strange meson fields $\sigma^{*}$ and $\phi$ are
described by Klein-Gordon and Proca equations, respectively,

\begin{eqnarray} \label{sigmamotion}
(\partial_{\mu}\partial^{\mu}+m_{\sigma}^{2})\sigma&=&
g_{\sigma}\Big[\sum_{B}\frac{g_{\sigma B
}}{g_{\sigma}}\rho_{B}^{S}
\nonumber\\
&&-b_{\sigma}M{(g_{\sigma}\sigma)}^{2}
\nonumber\\
&&-c_{\sigma}{(g_{\sigma}\sigma)}^{3}\Big]\;,
\\
\label{omegamotion}
\partial_{\mu}\omega^{\mu\nu}+m_{\omega}^{2}\omega^{\nu}&=&
g_{\omega}\Big[\sum_{B}\frac{g_{\omega B}}{g_{\omega}}\rho_{B}^{B}
\nonumber\\
&&-c_{\omega}g_{\omega}^{3} (\omega_{\mu}\omega^{\mu}\omega^{\nu})
\nonumber\\
&&-g_{\rho}^{2}\bm{\rho_{\mu}}.\bm{\rho}^{\mu}\Lambda_{V}
g_{\omega}\omega_{\mu}\Big]\;,
\\
\label{rhomotion}
\partial_{\mu}\bm{\rho}^{\mu\nu}+m_{\rho}^{2}\bm{\rho}^{\nu}&=&
g_{\rho}\Big[\sum_{B}\frac{g_{\rho
B}}{g_{\rho}}\rho_{B}^{B}\bm{\tau}
\nonumber\\
&&-g_{\rho}\bm{\rho_{\mu}}
\Lambda_{V}g_{\omega}^{2}\bm{\omega}_{\mu}\bm{\omega}^{\mu}\Big]\;,
\\
\label{deltamotion}
(\partial_{\mu}\partial^{\mu}+m_{\delta}^{2})\bm{\delta}&=&
g_{\delta}\sum_{B}\frac{g_{\delta B
}}{g_{\delta}}\rho_{B}^{S}\bm{\tau}\;,
\\
\label{sigmasmotion}
(\partial_{\mu}\partial^{\mu}+m_{\sigma^{*}}^{2})\sigma^{*}&=&g_{\sigma^{*}\Lambda}
\sum_{B}\frac{g_{\sigma^{*}B}}{g_{\sigma^{*}\Lambda}}\rho_{B}^{S}\;,
\\
\label{phimotion}
\partial_{\mu}\phi^{\mu\nu}+m_{\phi}^{2}\phi^{\nu}&=&g_{\phi\Lambda}\sum_{B}
\frac{g_{\phi B}}{g_{\phi\Lambda}}\rho_{B}^{B}\;.
\end{eqnarray}

This set of nonlinear equations is solved in the mean-field
approach, i.e., operators of meson fields are replaced by their
expectation values. Additionally, a no-sea approximation is
considered, which doesn't take account of the Dirac sea of
negative energy states.

Static, homogenous, infinite nuclear matter allows us to consider
some simplifications due to translational invariance and
rotational symmetry of the nuclear matter. This causes the
expectation values of spacelike components of vector fields vanish
and only zero components, $\rho_{0}$, $\omega_{0}$, and
$\phi_{0}$, remain. In addition, due to rotational invariance
around the third ax of isospin space only the third component of
isovector fields $\rho^{(3)}$ and $\delta^{(3)}$ survive. After
reducing of the equation of motion meson field potentials follow
directly:

\begin{eqnarray}
\label{sigmapotential} U_{\sigma}&\equiv&
-g_{\sigma}\bar{\sigma}=-\frac{g_{\sigma}^{2}}{m_{\sigma}^{2}}
[\sum_{B}\frac{g_{\sigma B}}{g_{\sigma}}\rho_{B}^{S}
\nonumber\\
&&-
b_{\sigma}M{(g_{\sigma}\bar{\sigma})}^{2}
-c_{\sigma}{(g_{\sigma}\bar{\sigma})}^{3}]
\;,
\\
\label{omegapotential} U_{\omega}&\equiv&
g_{\omega}\bar{\omega}_{0}=\frac{g_{\omega}^{2}}{m_{\omega}^{2}}
[\sum_{B}\frac{g_{\omega B }}{g_{\omega}}\rho_{B}^{B}
\nonumber\\
&&-c_{\omega}{(g_{\omega}\bar{\omega}_{0})}^{3}
-U_{\rho}^{2}\Lambda_{V}(g_{\omega}\bar{\omega}_{0})]\;,
\\
 \label{rhopotential} U_{\rho}&\equiv&
g_{\rho}\bar{\rho}_{0}^{(3)}= \frac{g_{\rho}^{2}}{m_{\rho}^{2}}
\bar{\psi}\gamma^{0}\tau_{3}\psi=
\nonumber\\
&&\frac{g_{\rho}^{2}}{m_{\rho}^{2}}[\sum_{B}\frac{g_{\rho B
}}{g_{\rho}}\rho_{B}^{B}\tau_{3 B}-g_{\rho}\bar{\rho}_{0}^{(3)}
\Lambda_{V}U_{\omega}^{2}]\;,
\end{eqnarray}
\begin{eqnarray}
\label{deltapotential} U_{\delta}&\equiv&
-g_{\delta}{\bar{\delta}}^{(3)}=
-\frac{g_{\delta}^{2}}{m_{\delta}^{2}}\sum_{B}\frac{g_{\delta
B}}{g_{\delta}}(\rho_{B}^{S}\tau_{3 B})\;,
\\
\label{sigmaspoteential} U_{\sigma^{*}}&\equiv& -g_{\sigma^{*}
\Lambda}\bar{\sigma^{*}}= -\frac{g_{\sigma^{*} \Lambda
}^{2}}{m_{\sigma^{*}}^{2}}\sum_{B}\frac{g_{\sigma^{*} B
}}{g_{\sigma^{*} \Lambda}}\rho_{B}^{S}\;,
\\
\label{phipotential} U_{\phi}&\equiv& g_{\phi
\Lambda}\bar{\phi}_{0}= \frac{g_{\phi \Lambda
}^{2}}{m_{\phi}^{2}}\sum_{B}\frac{g_{\phi B }}{g_{\phi
\Lambda}}\rho_{B}^{B}\;,
\end{eqnarray}
where $\tau_{3 p}=1, \tau_{3 n,\Sigma^{-}}=-1, \tau_{3
\Lambda^{0}}=0$ are isospin projections for baryons. Scalar
density $\rho_{S}$ is expressed as the sum of baryon
($B=p,n,\Lambda^{0},\Sigma^{-}$) contributions

\begin{equation} \label{densityscalar}
\rho_{B}^{S}=\frac{2J_{B}+1}{{(2\pi)}^{3}}\int_{0}^{k_{B}}d^{3}\!k
\, \frac{M_{B}^{*}}{{(\bm{k}^{2}+{M_{B}^{*}}^{2})}^{1/2}} \;.
\end{equation}
In Eq. (\ref{densityscalar}) $k_{B}$ is baryons' Fermi momentum,
$J_{B}$ corresponds to baryon spin and $M_{B}^{*}$ denotes baryon
effective masses which can be written as

\begin{equation}\label{nucleonmasses}
M_{B}^{*}=M-g_{\sigma B}\bar{\sigma}-g_{\delta B}
{\bar{\delta}}^{3}\tau_{3 B} -g_{\sigma^{*} B}\bar{\sigma^{*}}\;.
\end{equation}
Condensed scalar $\sigma$,$\delta$, and $\sigma^{*}$ meson fields
generate a shift of baryon masses, in consequence of which nuclear
matter is described as a system of pseudonucleons with masses
$M_{B}^{*}$ moving in classical fields. Note that while $\sigma$
meson shifts all the baryon masses (even with different strength
for nucleons and hyperons), $\delta$ meson field is responsible
for splitting of effective masses of baryons with nonzero isospin
only, which is an important feature of $\delta$ meson influence on
the nuclear matter saturation mechanism and its properties.
Additionally, due to no interaction of strange mesons with
nucleons, $\sigma^{*}$ meson shifts only hyperon masses.

Total baryon density is given by,

\begin{equation} \label{densitybaryon}
\rho_{B}=\sum_{B}\rho_{B}^{B}=\sum_{B}\frac{2J_{B}+1}{{(2\pi)}^{3}}\int_{0}^{k_{B}}d^{3}\!k
=\sum_{B}\frac{k_{B}^{3}}{3\pi^{2}} \, \;.
\end{equation}
The momentum-energy tensor

\begin{equation} \label{energytensor}
T_{\mu\nu}=-g_{\mu\nu}\mathcal{L}+\frac{\partial\Phi_{i}}{\partial
\bm{x}^{\nu}}
\frac{\partial\mathcal{L}}{\partial(\partial\Phi_{i}/\partial
\bm{x}_{\mu})}\;,
\end{equation}
where $\Phi_{i}$ generally denotes physical fields, gives the
energy density of the system as its zero component
$\varepsilon=\left\langle T_{00}\right\rangle$, and finally, the
binding energy per nucleon is related to the energy density by,

\begin{equation} \label{energypernucleon}
E_{b}=\frac{\varepsilon}{\rho_{B}}-M\;.
\end{equation}

While we are considering $\beta$-stable nuclear matter consisting
of baryons, hyperons, and leptons, there have to be fulfilled
equilibrium conditions of chemical potentials defined as Fermi
energy of particles at the top of Fermi sea, for baryons taking a
form,

\begin{equation}
\mu_{B}=g_{\omega B}\bar{\omega_{0}}+g_{\rho
B}\bar{\rho_{0}}^{(3)}\tau_{3 B}+g_{\phi
B}\bar{\phi_{0}}+\sqrt{k_{B}^{2}+{M_{B}^{*}}^{2}}\;,
\end{equation}
and for leptons,

\begin{equation}
\mu_{e,\mu}=\sqrt{k_{e,\mu}^{2}+{m_{e,\mu}}^{2}}\enspace.
\end{equation}
Generally, the chemical potential equilibrium conditions can be
written as,

\begin{equation}
\mu_{B}=q_{b,B}\mu_{n}-q_{e,B}\mu_{e}\;,
\end{equation}
where $q_{b,B}$ is baryon number of particles, and $q_{e,B}$
denotes electric charge of particles, giving a set of equilibrium
equations,

\begin{eqnarray}
\nonumber \mu_{\Lambda}&=&\mu_{n}\;,
\nonumber\\
\mu_{\Sigma^{-}}&=&\mu_{n}+\mu_{e}\;,
\nonumber\\
\mu_{p}&=&\mu_{n}-\mu_{e}\;,
\nonumber\\
\mu_{\mu}&=&\mu_{e}\;.
\end{eqnarray}
Simultaneously, charge neutrality:

\begin{equation} \label{sestdesiatjeden}
\rho_{p}=\rho_{e}+\rho_{\mu}+\rho_{\Sigma^{-}}\;,
\end{equation}
and the total baryon density (\ref{densitybaryon}) have to be met.

\section{Results and discussion}

The reference \cite{KBG03} deals with the role of $\delta$ meson
and vector cross interaction in asymmetric nuclear matter. The
present work extends this study to hyperon matter in
$\beta$-equilibrium.

Several different mean-field parametrizations are used, taken from
the Ref. \cite{KBG03}. The first pair MA and MB (listed in
Tab.~\ref{table1}) results from the fit of DBHF nuclear matter
calculations performed by Machleidt and co-workers \cite{BLM92}.
Binding energy per nucleon for symmetric matter and neutron
matter, and simultaneously symmetric matter isoscalar potentials
(all results for Bonn A NN potential) were taken into account.
They differ in degrees of freedom used - the second one does not
include the vector cross interaction term. Couplings of hyperons
to scalar meson were fitted to reproduce hypernuclei potential
well given by Eq. (\ref{hypwell}), and strengths of strange scalar
meson couplings are adjusted to fulfil potential depth from Eq.
(\ref{doublehypwell}). Similarly the second parametrization pair
LA and LB (also listed in Tab.~\ref{table1}) reproduces DBHF
results of Lee and co-workers \cite{LKL97} (also Bonn A
potential), where, in addition to the first fit, also binding
energies for several asymmetries were fitted together with
symmetric matter isoscalar potentials. The next parametrizations
HA - HD (which differ in inclusion or exclusion of cross
interaction and $\delta$ meson) ensue from the fit of DBHF results
of Huber and co-workers \cite{HWW95} (Bonn B potential), being
fitted to binding energy per nucleon for several asymmetries,
symmetric matter isoscalar potentials, and, unlike the previous
two sets, also to proton and neutron isoscalar potentials, which
enables an evaluation of the $\delta$ meson field contribution.
Isoscalar $\sigma$,$\omega$ mesons with their self-interactions,
isovector $\rho$ meson, $\rho$-$\omega$ cross interaction, and for
relevant parametrization sets also $\delta$ meson, were used as
degrees of freedom.

\begin{table} [h!]
\caption{\label{table1}Parameter sets resulting from the fit
\cite{KBG03} of Machleidt and co-workers \cite{BLM92} and Lee and
co-workers \cite{LKL97} DBHF results (for nucleon-meson coupling
constants). Hyperon-scalar meson couplings resulting from the fit
to hypernuclear potentials (\ref{hypwell}),(\ref{doublehypwell})
are also listed, hyperon-vector (isovector) meson couplings ensue
directly from Eqs. (\ref{quarkmodel}).}
\begin{ruledtabular}
\begin{tabular}{ccccc}

 & MA & MB & LA & LB \\

\hline

$g_{\sigma}^{2}$ & 106.85 & 112.27 & 103.91 & 102.11 \\

$g_{\omega}^{2}$ & 180.61 & 204.36 & 147.84 & 146.73 \\

$g_{\rho}^{2}$ & 18.445 & 9.493 & 17.432 & 9.670 \\

$b_{\sigma}$ & -0.00258 & -0.00298 & 0.000972 & 0.000836 \\

$c_{\sigma}$ & 0.0115 & 0.0133 & 0.00127 & 0.00124 \\

$c_{\omega}$ & 0.0158 & 0.0204 & 0.00542 & 0.00519 \\

$\Lambda_{V}$ & 0.2586 &     & 0.1879 &     \\

$\chi^{2}/N$  & 2.76  & 9.95 & 1.69 & 2.62 \\

\hline

$g_{\sigma\Lambda}/g_{\sigma N}$& 0.602 & 0.599 & 0.602 & 0.601 \\

$g_{\sigma\Sigma}/g_{\sigma N}$& 0.472 & 0.473 & 0.475 & 0.476 \\

$g_{\sigma^{*}\Lambda(\Sigma)}/g_{\sigma N}$& 0.646 &
0.662 & 0.603 & 0.605 \\

\end{tabular}
\end{ruledtabular}
\end{table}

\begin{table} [h!]
\caption{\label{table2}Parameter sets resulting from the fit
\cite{KBG03} to Huber and co-workers DBHF results \cite{HWW95}.
Hyperon couplings are obtained analogically as in
Tab.~\ref{table1}.}

\begin{ruledtabular}
\begin{tabular}{ccccc}

& HA & HB & HC & HD \\

\hline

$g_{\sigma}^{2}$ & 90.532  & 86.432  & 91.110 & 87.591 \\

$g_{\omega}^{2}$ & 108.95  & 106.89  & 109.26 & 107.61 \\

$g_{\rho}^{2}$ & 36.681 & 28.795 & 20.804 & 15.335 \\

$g_{\delta^{2}}$ &  28.739 & 25.170 &     &     \\

$b_{\sigma}$ & 0.00439  & 0.00338 & 0.00444 & 0.00357 \\

$c_{\sigma}$ & -0.00520 & -0.00378 & -0.00521 & -0.00398 \\

$c_{\omega}$ & -0.000142 & -0.00105 & -0.0000385 & -0.000775 \\

$\Lambda_{V}$ & 0.1065 &       & 0.3481 &     \\

$\chi^{2}/N$  & 2.05 & 3.80 & 5.85 & 6.89 \\

\hline

$g_{\sigma\Lambda}/g_{\sigma N}$& 0.605 & 0.600 & 0.606 & 0.601 \\

$g_{\sigma\Sigma}/g_{\sigma N}$& 0.446 & 0.449 & 0.446 & 0.448 \\

$g_{\sigma^{*}\Lambda(\Sigma)}/g_{\sigma N}$& 0.576 &
0.582 & 0.576 & 0.580 \\

\end{tabular}
\end{ruledtabular}
\end{table}

By simultaneous calculation of the equilibrium conditions and
meson field potentials the dependence of particle fractions on
baryon density was obtained. We considered $\beta$-stable nuclear
matter consisting of protons, neutrons, $\Sigma^{-}$,
$\Lambda^{0}$ hyperons, electrons and muons. We did not take into
account other types of hyperons because the heaviest
$\Delta^{\frac{+}{0},++}$ seem to be suppressed \cite{HWW98,Gl85}
and also other baryons
($\Sigma^{{\frac{0}{}}}$,$\Xi^{\frac{0}{}}$) due to their electric
charge or higher rest mass may appear only at very high densities,
or may not appear at all \cite{VPR00}. Since the parametrizations
from Ref. \cite{KBG03} result from fits up to densities around
0.33 fm$^{-3}$, the reliability of the extrapolation to such a
high density region decreases. Moreover, it seems that at least
some important matter properties are more affected by total
hyperonization than by hyperon subtypes themselves, which is
indicated for example by very similar influence of hyperons on
electron chemical potential with presence or with total exclusion
of $\Sigma^{-}$ (due to its possible strong repulsion in nuclear
medium), which can be compensated by $\Xi^{-}$ \cite{Gl01}.
Therefore, the particular abundance of baryon species is not very
important \cite{BLC99}.

\begin{figure}[t!]
\includegraphics[width =8cm]{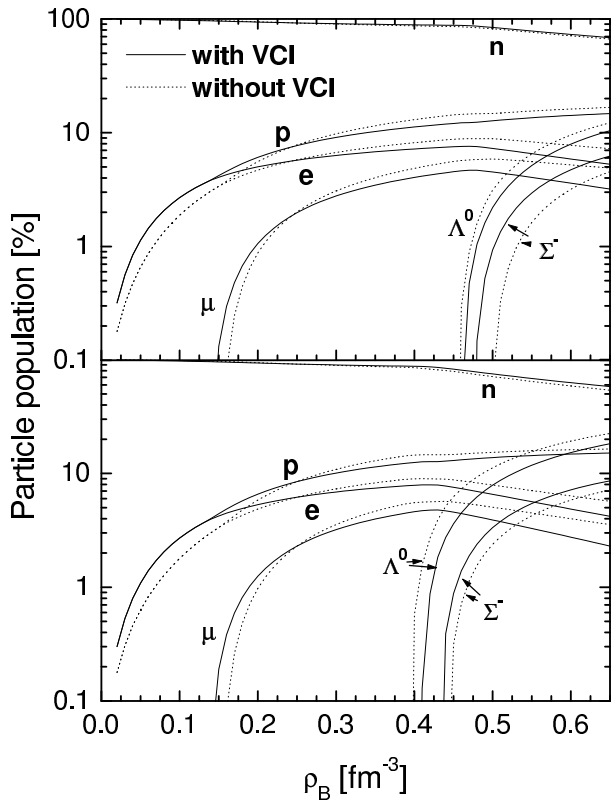}
\caption{\label{PopulMaLee} Populations of particles in
$\beta$-stable hyperon matter, calculated with parametrization
sets which reproduce Dirac-Brueckner-Hartree-Fock results obtained
by Machleidt {\sl et al.} \cite{BLM92} [upper panel], and that of
Lee {\sl et al.} \cite{LKL97} [lower panel]. Solid lines
correspond to the fit and equilibrium calculations performed with
vector cross interactions (VCI) (MA, LA), dotted lines denotes
calculations where cross interaction was excluded (MB, LB).
Nucleon-meson parametrizations are taken from Ref. \cite{KBG03}
and listed in Tab.~\ref{table1}.}
\end{figure}

In Fig.~\ref{PopulMaLee} the results for parametrizations MA (MB)
(upper panel) and LA (LB) (lower panel) are shown. Starting from
almost pure neutron matter with rising total baryon density the
nuclear matter is enriched by protons and, consequently, due to
charge neutrality also by electrons. Subsequently, the electron
chemical potential is rising and when it reaches value of the rest
mass of muon it become possible to turn electron into muon. This
happens already closely below the saturation density at
$\rho_{B}=0.14$ fm$^{-3}$ and causes also slight growth of proton
fraction rise. With further rise of density it turns up
energetically favorable to turn nucleons with high kinetic energy
into the heavier baryon resonances. The first hyperon appearing is
the $\Lambda^{0}$ at density 0.46 fm$^{-3}$.

In many older works (e.g., \cite{VPR00,Gl01,HKL02,SM96,BLC99}) it
is $\Sigma^{-}$ hyperon that appear as the first one. This was due
to its negative electric charge and zero isospin of the lightest
$\Lambda^{0}$ -- the electron chemical potential together with
$\rho$ meson field affecting $\Sigma^{-}$ was higher than
$\Sigma^{-}$ and $\Lambda^{0}$ effective mass difference. However,
this was in the case of assuming the negative potential well which
feels $\Sigma^{-}$ in nuclear matter, whereas later investigation
shows that it feels more probably a strong repulsion. Thus, the
$\Lambda^{0}$ onset is followed by $\Sigma^{-}$ hyperon at density
0.48 fm$^{-3}$. Nevertheless, we have to note that quantitative
value of the threshold densities depends strongly on poorly known
nucleon-hyperon coupling constants and their better knowledge will
improve correspondence of theory with reality. In any case, the
onset of hyperons has distinct influence on nucleon and lepton
populations. Hyperons due to the chemical potential relations
heighten proton fraction following by a fall of number of both
leptons in order to hold the charge neutrality of matter. The
lepton population reduction is a consequence of the local
non-conservation of lepton number (neutrinos escape from matter).
These general characteristics of density dependence of
$\beta$-stable hyperon matter are (except the $\Sigma^{-}$
repulsion) qualitatively in accordance with the above mentioned
works.

For examination of the vector cross interactions the results for
parametrization without them (MB) are drawn in the same plot by
dotted lines. Since VCI are softening the equation of state of
nuclear matter, it generally enforces neutron population to the
prejudice of protons with slight shift of negatively charged muons
and $\Sigma^{-}$ onset to the lower densities and that of neutral
$\Lambda^{0}$ to the higher densities, thus reducing the onset
density difference. Without the VCI muons appear at 0.16
fm$^{-3}$, $\Lambda^{0}$ at 0.45 fm$^{-3}$ and $\Sigma^{-}$ at
0.50 fm$^{-3}$.

The lower panel of Fig.~\ref{PopulMaLee} shows populations for the
second parametrization pair LA (LB) where the muons appear at 0.14
fm$^{-3}$ (0.16 fm$^{-3}$), $\Lambda^{0}$ at 0.41 fm$^{-3}$ (0.40
fm$^{-3}$), and $\Sigma^{-}$ at 0.43 fm$^{-3}$ (0.44 fm$^{-3}$),
values in the brackets being for non-VCI case. The main difference
in comparison with the previous parameter set is lowering the
hyperon thresholds and generally richer hyperonization of matter.
While in MA (MB) set the total hyperonization of matter at density
$\rho_{B}=0.6$ fm$^{-3}$ is 12.7\% (12.7\%), in LA (LB) results it
is 22.4\% (25.1\%). This is a natural consequence of harder EOS in
the second parametrization set.

\begin{figure}[t!]
\includegraphics[width =8cm]{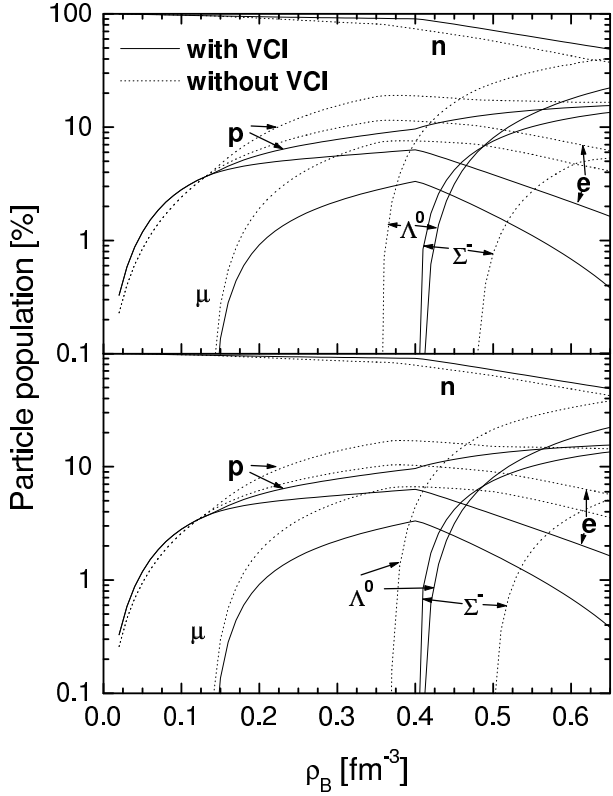}
\caption{\label{PopulHub1} Particle populations calculated
similarly as in Fig.~\ref{PopulMaLee} with nucleon-meson
parametrization sets taken from Ref. \cite{KBG03}, here relisted
in Tab.~\ref{table2}, which reproduces DBHF-calculated asymmetric
nuclear matter properties performed by Huber {\sl et al.}
\cite{HWW95}. In both panels solid lines represent results of fit
equilibrium calculations performed with $\delta$ meson as well as
vector cross interaction (HA). Dotted lines in the upper panel are
results with $\delta$ meson inclusion but without VCI (HB),
whereas dotted lines in the lower panel represents populations
with parametrization which does account of neither $\delta$ meson
nor VCI (HD).}
\end{figure}

The next parametrizations HA-D give outcome plotted in
Fig.~\ref{PopulHub1}. Solid line in both panels denotes particle
population density dependence with inclusion of both $\delta$
meson and VCI (HA). The influence of the VCI absence with
simultaneous $\delta$ meson inclusion is seen from dotted lines in
the upper panel (HB) while an analogical situation in $\delta$
meson absence can be seen from dotted lines in the lower panel
(HD). The VCI effect is more distinct than in the previous two
parameter sets. First, in density range roughly from one to two
times saturation density there is a strong shortage of protons
when VCI is present, independently on the $\delta$ meson. Second
important point is a shift of $\Lambda^{0}$ threshold to higher
densities from 0.36 fm$^{-3}$ to 0.41 fm$^{-3}$ with $\delta$
meson or from 0.37 fm$^{-3}$ to 0.41 fm$^{-3}$ without it as well
as a shift of $\Sigma^{-}$ threshold to lower densities. The
influence of VCI on hyperon thresholds is in this case so strong
that in spite of the strong repulsion of $\Sigma^{-}$ hyperon in
nuclear matter it appears as the first hyperonic species.
Consequently, the third effect of VCI is more rapid
deleptonization of matter -- at density 0.6 fm$^{-3}$ there are
0.69\% (4.85\%) of muons and 2.19\% (7.28\%) of electrons in HA
(HB) case, while 2.75\% (4.00\%) of muons and 4.87\% (6.34\%) of
electrons in LA (LB) parametrizations, and 3.60\% (5.22\%) of
muons and 5.88\% (7.75\%) of electrons in MA (MB) parameter sets.
It is clear that the VCI play primary role in equilibrated matter
and it is an indispensable physical degree of freedom; this
supports conclusions from \cite{KBG03}.

\begin{figure}[t!]
\includegraphics[width =8cm]{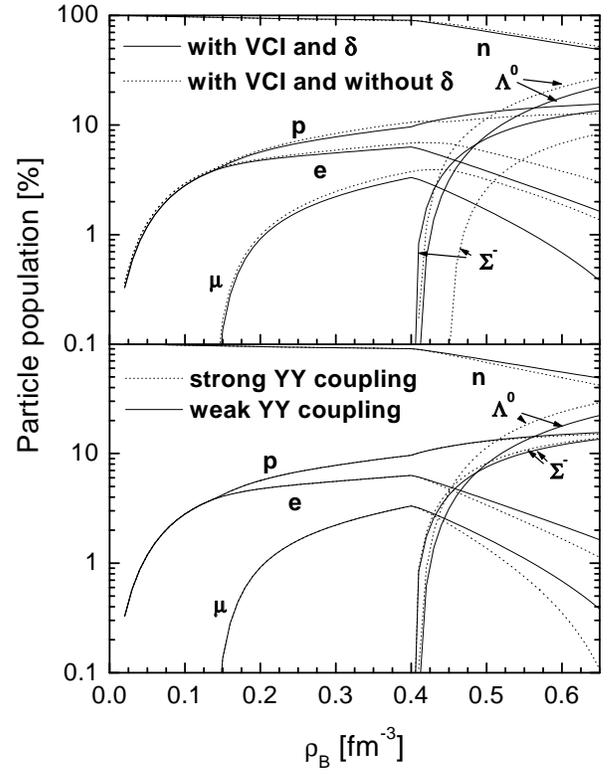}
\caption{\label{PopulHub2} Influence of isovector $\delta$ meson
[upper panel] and strange mesons [lower panel] on the particle
populations. Solid lines represent calculations with both $\delta$
and strange mesons (HA) with weak YY interactions, dotted lines in
the upper panel are populations without $\delta$ meson (HC), and
dotted lines in the lower panel denote the results where YY
interactions were strengthened ($U_{\Lambda}^{(\Lambda)}= -20$
MeV).}
\end{figure}

To evaluate independently an effect of the isoscalar scalar degree
of freedom the particle populations are drawn again in
Fig.~\ref{PopulHub2}, upper panel, both VCI including and
explicitly with (solid lines, HA) and without (dotted ones, HC)
$\delta$ meson. It influences mostly the onset of $\Sigma^{-}$
hyperon (isovector $\delta$ meson does not interact with
$\Lambda^{0}$), however, not to such amount as VCI. Moreover, the
total hyperonization is also affected only slightly - presence of
$\delta$ meson increases it from 29.0 \% only to 29.8 \%. One can
conclude that in spite of its isospin nature the role of $\delta$
meson is of a secondary importance also in highly asymmetric
matter. This is in agreement with conclusions of Ref. \cite{SM96}.

\begin{figure}[t!]
\includegraphics[width =8cm]{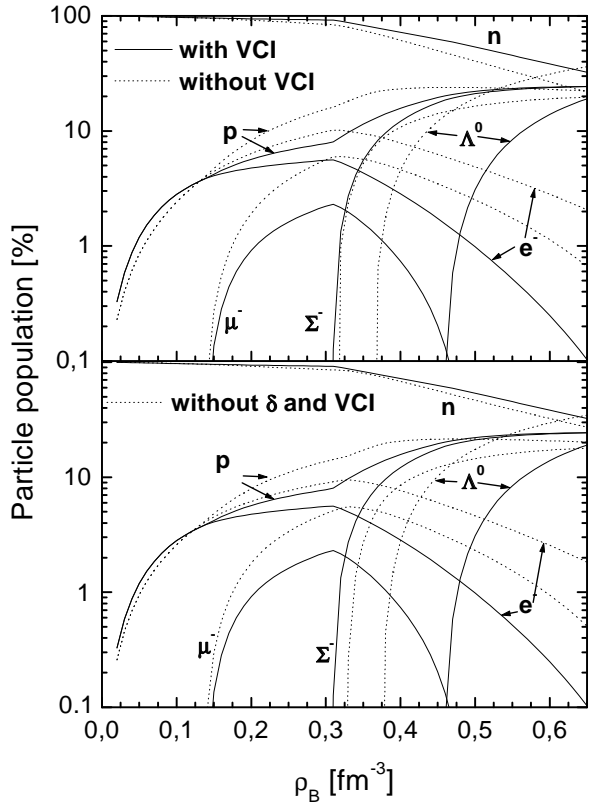}
\caption{\label{PopulHub3} Particle populations calculated using
attractive potential for $\Sigma^{-}$ hyperon in nuclear matter:
$U_{\Sigma^{-}}^{(N)}= -30$ MeV. Analogically to
Fig.~\ref{PopulHub1}, in both panels solid lines represent results
of fit equilibrium calculations performed with $\delta$ meson as
well as vector cross interaction (HA). Dotted lines in the upper
panel are results with $\delta$ meson inclusion but without VCI
(HB), whereas dotted lines in the lower panel represent
populations with parametrization which does account of neither
$\delta$ meson nor VCI (HD).}
\end{figure}

In the same figure (lower panel) the role of the strength of
hyperon-hyperon interaction mediated by strange $\sigma^{*}$ and
$\phi$ mesons is examined. Solid lines represent results employing
weak YY interaction (Eq. (\ref{doublehypwell})) while dotted lines
correspond to enforced YY coupling with potential well of
$\Lambda$ hyperon in hyperonic matter U$_{\Lambda}^{(\Lambda)}=
-20$ MeV. As strange mesons are not coupled to nucleons, their
effect starts up above the $\Sigma^{-}$ threshold density (which
is not affected) and is only of a small magnitude. Another effect
of strange mesons is a stabilization of matter at very high
densities. For LA (B) negative nucleon effective masses appear
above 1.89 fm$^{-3}$ (1.76 fm$^{-3}$) which are changed to a
positive value if $\sigma^{*}$, $\phi$ are introduced. Even there
does not occur also negative electron chemical potential, this is
in accordance with results of Ref. \cite{SM96}.

Results of calculations in the case of $\Sigma^{-}$ hyperon
feeling an attractive interaction (negative potential well
U$_{\Sigma^{-}}^{(N)}= -30$ MeV) which seemed to be realistic in
older works, are drawn in Fig.~\ref{PopulHub3}. Analogically to
Fig.~\ref{PopulHub1}, on both panels there are shown results for
HA parametrization (with hyperon couplings modified to reproduce
negative potential well of $\Sigma^{-}$) with solid lines. In the
upper panel the dotted lines represent results from the
parametrization HB (without VCI), and the dotted lines in the
lower panel correspond to parametrization HD (without VCI and
$\delta$ meson). We can see that in spite of strong (and natural)
decrease of threshold baryon density for $\Sigma^{-}$ hyperon
onset, the general features of VCI influence are the same. Also in
this case VCI enhance the deleptonization process, lowers total
hyperonization of matter and strongly affect population of protons
in the middle-density region.

\begin{figure}[t!]
\includegraphics[width =8cm]{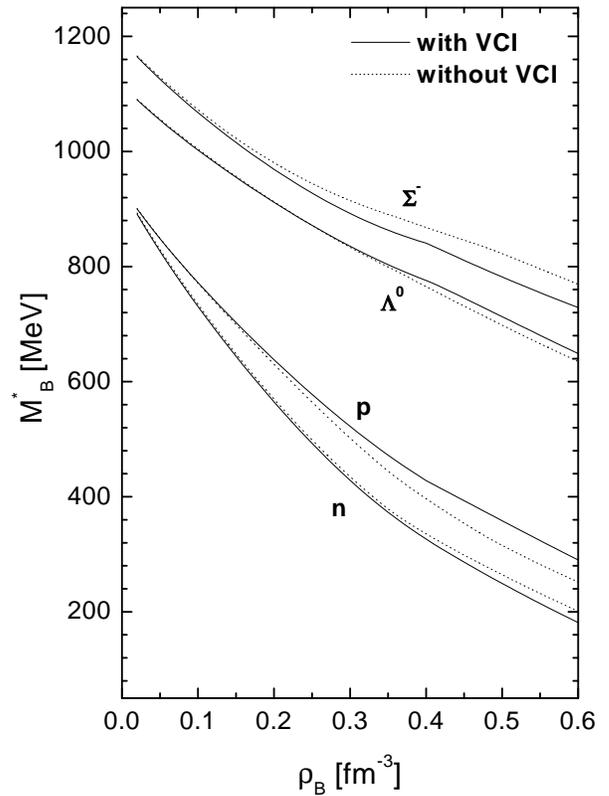}
\caption{\label{Masses} Density dependence of the effective baryon
masses in $\beta$-stable hyperon matter calculated with
parametrization including $\delta$ meson, vector cross interaction
and strange mesons, listed in Tab.~\ref{table2} (solid lines, HA),
and without VCI (dotted lines, HB).}
\end{figure}

\begin{figure}[t!]
\includegraphics[width =8cm]{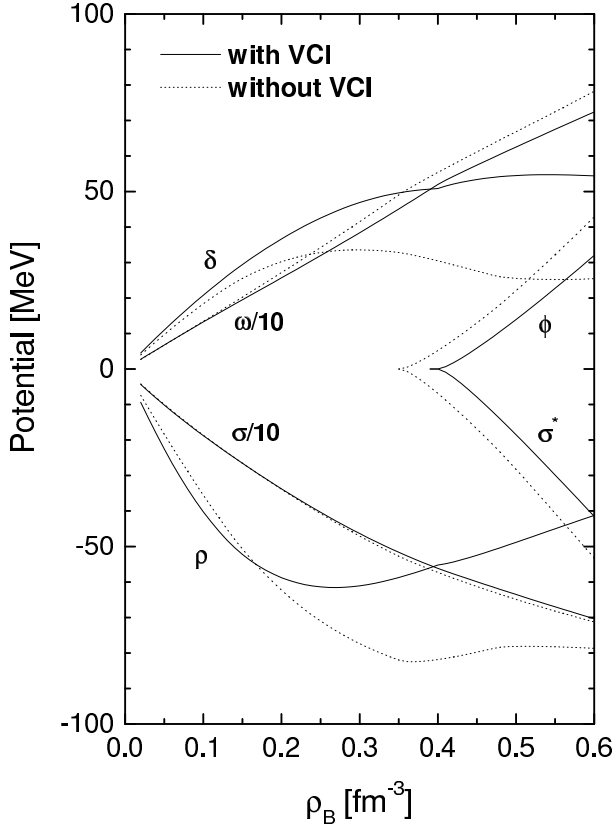}
\caption{\label{Potentials} Meson field potentials calculated with
the same full parametrization set as in the previous figure (solid
line, HA), and with exclusion of vector cross interaction (dotted
line, HB). For practical purposes isoscalar $\sigma$ and $\omega$
fields are shown divided by factor 10.}
\end{figure}

Besides not very strong effect of $\delta$ meson on the
composition of $\beta$-stable matter, due to its isospin vector
nature it has an impact on effective masses of baryons with
non-zero isospin. It can be seen from Fig.~\ref{Masses} where
nucleon masses are splitted. Nonetheless, in the VCI presence the
$\delta$ potential saturation at about 55 MeV occurs after hyperon
appearance, as it flows from Fig.~\ref{Potentials}. This
saturation is given by a balance of the total baryon density and
by fractions of non-zero isospin baryons and additionally supports
our conclusion and conclusion from Ref. \cite{SM96} of small
$\delta$ meson influence. Lower $\delta$ meson field in VCI
absence (saturation at 30 MeV) is given partially by lower
coupling of $\delta$ meson to nucleons
($g_{\delta}^{2}/g_{\delta\textrm{VCI}}^{2}=0.88$) needed for
proper reproduction of DBHF calculations as well as by richer
proton and $\Lambda^{0}$ and poorer neutron and $\Sigma^{-}$
populations, thus resulting in more isospin symmetric matter. Even
stronger saturation mechanism occurs for the $\rho$ meson field,
in this case it is a direct impact of cross interactions with
$\omega$ field which also exhibits decrease, though due to much
weaker $\rho$ field not leading to a saturation. Finally, since
the VCI soften EOS, in their absence there is a higher
hyperonization of matter -- including (excluding) the
isovector-scalar field it is 40.7 \% (36.8 \%), compared to 29.8
\% (29.0 \%) at $\rho_{B}=0.6$ fm$^{-3}$ if they are switched on,
when weaker strange fields are also another natural consequence.

\begin{figure}[t!]
\includegraphics[width=8cm]{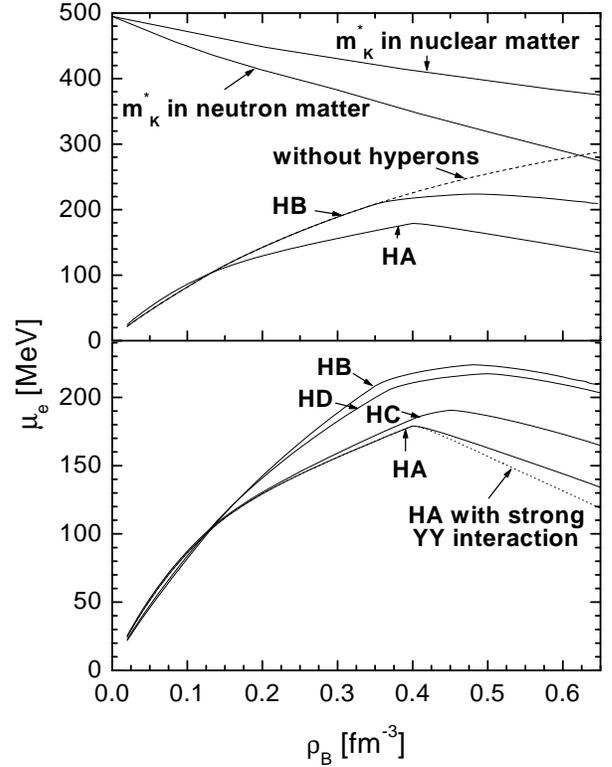}
\caption{\label{Chempot} Chemical potentials of electrons in
$\beta$-stable hyperon matter obtained with parametrization with
$\sigma$,$\omega$,$\rho$,$\delta$ mesons, isoscalar
self-interactions, and vector cross interaction
(Tab.~\ref{table2}, HA) and without cross interactions (HB), as
well as density dependence of the electron chemical potential in
matter without hyperons (dotted line). For comparison, the
effective mass of $K^{-}$ in nuclear matter (taken from
\cite{MPH99}) and neutron matter \cite{WRW97} is shown. The lower
panel shows the results for remaining parametrizations HC and HD
and, additionally, the results of HA but without strange mesons.}
\end{figure}

Another important feature of the hyperon onset is a strong change
in the density dependence of electron chemical potential. Distinct
decline of the electron Fermi momentum and concentration in matter
after hyperonization is immediately followed by a drop of chemical
potential which has serious consequences for an occurance of kaon
condensation, as was first pointed out in Ref. \cite{Gl85}.
Density dependence of the chemical potential of electrons in
equilibrated matter for parametrizations examined in this paper
are plotted in Fig.~\ref{Chempot}. The upper panel shows chemical
potential resulting same as in the previous figures from
parametrization with $\sigma$,$\omega$,$\rho$, and $\delta$ mesons
with (HA) and without VCI (HB). The recent effective kaon mass
resulting for nuclear matter \cite{MPH99} and neutron matter
\cite{WRW97} is also plotted. However, its value is still an
actual topic of research and at least for neutron matter may be
even lower (around 200 MeV at three times of saturation density),
as suggest recent analysis \cite{LLB97} of experimental data
\cite{Ba97}. If the electron chemical potential reaches effective
kaon mass the kaon condensation occurs. Nevertheless, as could be
expected, the electron chemical potential stops to rise after
hyperonization (otherwise it would continually increase as shows
dashed line) and starts to decline with increasing density.

The vector cross interaction affects both the maximum value of
chemical potential and the slope of its decrease. The VCI also
causes an immediate reduction of the chemical potential, while in
their absence there is a mild rise after hyperon onset as can be
better seen from the lower panel. There is more detailed
comparison of different parameter sets. Without VCI, the electron
chemical potential saturates at value about 224 MeV and $\delta$
meson has only a weak effect (about 6 MeV), otherwise the
saturation value is around 191 MeV and even 179 MeV with $\delta$
meson. Due to almost zero strange meson fields at break-point the
stronger YY coupling has no influence on the saturation value and
for higher densities it mildly fortifies VCI influence on the
electron chemical potential. Additionally, we have performed
analysis of the situation if $\Sigma^{-}$ hyperon feels attractive
potential in hyperon matter (or negative potential well) and it is
important to note that in this case the effect of VCI is even
stronger - it brings additional decrease of electron chemical
potential approximately by 30 \% (for clarity these results are
not shown in the Fig.~\ref{Chempot}). If $m_{K^{-}}=200$ MeV at
$\rho=3\rho_{0}$ verges on reality, it might not be clear if
inclusion of hyperons themselves would prevent kaon condensation.
Nevertheless, considering of VCI could additionally shift it to
much higher densities, possibly already not relevant for neutron
stars. Thus, VCI strongly supports hyperons in their role of kaon
condensation reduction factor, although intensity of this effect
depends on the exact kaon effective mass in nuclear medium as well
as accurate hyperon-nucleon constants.

\section{Summary}

In this work, several parametrizations of the relativistic
mean-field theory reproducing three different
Dirac-Brueckner-Hartree-Fock calculations for asymmetric nuclear
matter are used to calculate properties of hyperon matter in
$\beta$-equilibrium. Besides the isoscalar meson with their
self-interactions and isovector $\rho$ meson, also
isovector-scalar $\delta$ meson, and scalar-vector cross
interactions are considered and their influence on composition of
equilibrated hyperon matter and the electron chemical potential is
studied. Supporting previous conclusions the results obtained
indicate that VCI is an important degree of freedom with the
distinct impact on both the composition of matter and the electron
chemical potential. Since they soften nuclear equation of state,
especially if the $\delta$ meson is present, they reduce
hyperonization of matter and heighten population of neutrons, thus
making the neutron star matter more neutron-rich. Notwithstanding,
they simultaneously strongly support hyperons in their role of
kaon condensation reduction factor, resulting in a lower
saturation value of the electron chemical potential and also its
more steep decreasing after the hyperon onset. That shifts the
kaon condensation appearance to even higher densities and makes
the neutron star matter more neutron-rich also from this point of
view. It is interesting that neither stronger YY interaction nor
attractive potential of $\Sigma^{-}$ hyperon lessen this
influence.

The next step is to apply this parametrizations also to neutron
star properties calculation. Such a work is in progress.

\begin{acknowledgments}
This work was supported in part by the Slovak Grant Agency for
Science VEGA under Grant No. 2/4098/04. Calculations were
performed at the Computational Centre of Slovak Academy of
Sciences.
\end{acknowledgments}

\appendix*
\section{}

\begin{table*} [h!]
\caption{\label{Table3} Dependence of the binding energy per
nucleon $E_{b}$ (MeV) and the nuclear pressure $p$ (MeV.fm$^{-3}$)
on the baryon density $\rho_{B}$ (fm$^{-3}$) in
$\beta$-equilibrated matter for all of the parameterizations
considered in this article. The onset of hyperons causes a
significant change especially in the nuclear pressure behavior.}

\begin{ruledtabular}
\begin{tabular}{ccccccccccccccccc}

\multicolumn {1}{c}{} & \multicolumn {2}{c}{MA} & \multicolumn
{2}{c}{MB} & \multicolumn {2}{c}{LA} & \multicolumn {2}{c}{LB} &
\multicolumn {2}{c}{HA} & \multicolumn {2}{c}{HB} & \multicolumn
{2}{c}{HC} & \multicolumn {2}{c}{HD} \\

$\rho_{B}$ & $E_{b}$ & $p$ & $E_{b}$ & $p$ & $E_{b}$ & $p$ &
$E_{b}$ & $p$ & $E_{b}$ & $p$ &
$E_{b}$ & $p$ & $E_{b}$ & $p$ & $E_{b}$ & $p$ \\

\hline

0.02 & 6.28 & 0.06 & 6.45 & 0.06 & 3.87 & 0.03 & 3.15 & 0.02
& 3.03 & 0.02 & 2.85 & 0.01 & 3.14 & 0.02 & 2.93 & 0.02 \\

0.04 & 8.46 & 0.08 & 7.92 & 0.04 & 5.09 & 0.09 & 3.68 & 0.04 &
4.02 & 0.08 & 3.45 & 0.05 & 4.19 & 0.09 & 3.62 &
0.06 \\

0.06 & 8.64 & 0.06 & 8.05 & -0.02 & 6.23 & 0.21 & 4.24 & 0.13 &
5.12 & 0.23 & 4.29 & 0.19 & 5.38 & 0.23 & 4.54 &
0.20 \\

0.08 & 8.94 & 0.12 & 8.03 & 0.05 & 7.49 & 0.34 & 5.11 & 0.34 &
6.56 & 0.48 & 5.57 & 0.49 & 6.76 & 0.47 & 5.86 &
0.49 \\

0.1 & 9.46 & 0.36 & 8.42 & 0.33 & 8.94 & 0.78 & 6.36 & 0.73 & 8.17
& 0.83 & 7.31 & 0.99 & 8.31 & 0.81 & 7.59 &
0.98 \\

0.12 & 10.39 & 0.85 & 9.37 & 0.89 & 10.62 & 1.3 & 8.03 & 1.35 &
9.88 & 1.28 & 9.52 & 1.76 & 10.00 & 1.27 & 9.75 &
1.69 \\

0.14 & 11.81 & 1.61 & 10.89 & 1.75 & 12.55 & 2.02 & 10.11 & 2.24 &
11.72 & 1.86 & 12.18 & 2.82 & 11.85 & 1.88 & 12.28
& 2.66 \\

0.16 & 13.65 & 2.61 & 12.9 & 2.87 & 14.73 & 2.96 & 12.58 & 3.42 &
13.68 & 2.58 & 15.25 & 4.17 & 13.84 & 2.66 & 15.15 &
3.88 \\

0.18 & 15.87 & 3.88 & 15.32 & 4.21 & 17.17 & 4.16 & 15.42 & 4.88 &
15.76 & 3.49 & 18.68 & 5.86 & 15.99 & 3.65 & 18.32
& 5.38 \\

0.2 & 18.42 & 5.40 & 18.07 & 5.79 & 19.87 & 5.69 & 18.59 & 6.66 &
17.99 & 4.67 & 22.47 & 7.39 & 18.35 & 4.93 & 21.79
& 7.23 \\

0.22 & 21.26 & 7.18 & 21.09 & 7.62 & 22.85 & 7.56 & 22.07 & 8.79 &
20.43 & 6.18 & 26.61 & 10.47 & 20.93 & 6.55 & 25.54
& 9.47 \\

0.24 & 24.34 & 9.22 & 24.35 & 9.67 & 26.12 & 9.82 & 25.85 & 11.33
& 23.11 & 8.13 & 31.12 & 13.52 & 23.77 & 8.58 & 29.61
& 12.17 \\

0.26 & 27.65 & 11.49 & 27.79 & 11.94 & 29.66 & 12.48 & 29.92 &
14.26 & 26.08 & 10.58 & 35.99 & 17.15 & 26.89 &
11.06 & 33.99 & 15.39 \\

0.28 & 31.13 & 14.02 & 31.4 & 14.44 & 33.49 & 15.56 & 34.27 &
17.61 & 29.37 & 13.64 & 41.26 & 21.44 & 30.31 & 14.07 &
38.72 & 19.22 \\

0.3 & 34.79 & 16.78 & 35.15 & 17.15 & 37.59 & 19.06 & 38.89 &
21.40 & 33.04 & 17.39 & 46.93 & 26.48 & 34.07 & 17.96 &
43.8 & 23.75 \\

0.32 & 38.58 & 19.77 & 39.01 & 20.08 & 41.95 & 22.99 & 43.77 &
25.63 & 37.11 & 21.94 & 53.03 & 32.36 & 38.17 & 21.98 &
49.27 & 29.05 \\

0.34 & 42.49 & 22.98 & 42.98 & 23.21 & 46.56 & 27.36 & 48.89 &
30.29 & 41.60 & 27.28 & 59.57 & 39.16 & 42.65 &
27.01 & 55.14 & 35.24 \\

0.36 & 46.52 & 26.42 & 47.03 & 26.54 & 51.41 & 32.17 & 54.24 &
35.39 & 46.55 & 33.55 & 68.86 & 101.49 & 47.51 &
32.84 & 61.45 & 42.39 \\

0.38 & 50.64 & 30.07 & 51.16 & 30.08 & 56.47 & 37.36 & 59.8 &
40.93 & 51.95 & 40.74 & 86.73 & 140.21 & 52.78 & 39.57
& 72.54 & 116.54 \\

0.4 & 54.84 & 33.92 & 55.36 & 33.81 & 61.74 & 42.96 & 67.46 &
93.98 & 57.83 & 49.03 & 106.59 & 162.80 & 58.46 & 47.18 &
90.09 & 148.49 \\

0.42 & 59.12 & 37.99 & 59.61 & 37.73 & 69.99 & 101.19 & 80.49 &
122.43 & 71.78 & 143.46 & 127.21 & 183.38 & 68.62 &
125.64 & 109.16 & 171.86 \\

0.44 & 63.45 & 42.24 & 63.91 & 41.85 & 83.12 & 142.38 & 94.76 &
139.07 & 89.30 & 174.23 & 148.15 & 203.10 & 84.12 &
162.03 & 128.88 & 192.78 \\

0.46 & 67.99 & 67.93 & 69.28 & 82.93 & 98.09 & 159.37 & 110.61 &
169.26 & 107.49 & 193.50 & 169.23 & 223.95 &
101.86 & 198.73 & 148.89 & 212.69 \\

0.48 & 76.46 & 121.59 & 78.02 & 107.69 & 113.09 & 172.26 & 126.53
& 181.79 & 125.86 & 211.5 & 190.51 & 261.89 & 120.79
& 218.90 & 169.04 & 232.37 \\

0.5 & 87.46 & 140.35 & 87.76 & 135.01 & 127.86 & 182.89 & 142.14 &
192.69 & 144.18 & 228.49 & 212.54 & 287.25 &
139.77 & 236.53 & 189.30 & 261.08 \\

0.52 & 98.72 & 152.09 & 98.83 & 151.26 & 142.28 & 192.86 & 157.35
& 202.69 & 162.38 & 245.77 & 234.58 & 308.91 &
158.61 & 253.38 & 210.36 & 285.31 \\

0.54 & 109.90 & 161.65 & 109.98 & 161.52 & 156.31 & 201.47 &
172.12 & 212.12 & 180.39 & 261.12 & 256.51 & 327.81 &
177.22 & 269.57 & 231.41 & 305.71 \\

0.56 & 120.88 & 169.94 & 120.96 & 170.14 & 169.93 & 210.46 &
186.45 & 221.06 & 198.21 & 277.13 & 278.29 & 346.86 &
195.59 & 285.65 & 252.29 & 326.19 \\

0.58 & 131.58 & 177.62 & 131.67 & 177.76 & 183.14 & 218.05 &
200.33 & 229.80 & 215.76 & 293.46 & 299.96 & 368.37 &
213.67 & 301.69 & 272.99 & 346.33 \\

0.6 & 142.00 & 184.59 & 142.09 & 184.71 & 195.94 & 224.73 & 213.79
& 238.28 & 233.15 & 309.79 & 321.51 & 388.65 & 231.46
& 319.27 & 293.48 & 367.04 \\

0.62 & 152.12 & 191.33 & 152.21 & 191.34 & 208.35 & 230.68 &
226.83 & 246.52 & 250.27 & 326.63 & 345.63 & 410.32 & 248.96
& 333.33 & 313.79 & 388.07 \\

0.64 & 161.93 & 197.69 & 162.01 & 197.51 & 220.38 & 236.64 &
239.46 & 254.68 & 267.15 & 344.56 & 368.37 & 431.83 &
266.17 & 349.28 & 333.92 & 410.47 \\

\end{tabular}
\end{ruledtabular}
\end{table*}

\end{document}